\documentclass[final,5p,times,twocolumn]{elsarticle}

\usepackage{amssymb}
\usepackage{amsmath}

\usepackage{xurl} 
\usepackage{float}
\usepackage{makecell}
\usepackage{pgf-umlsd} 

\makeatletter
\def\ps@pprintTitle{%
  \let\@oddhead\@empty \let\@evenhead\@empty
  \let\@oddfoot\@empty \let\@evenfoot\@empty}
\makeatother
\begin{document}

\begin{frontmatter}



\title{Identity-Bound Academic Credentials on Blockchain:\\ On-Chain Issuer Accreditation with ERC-3643 and OnchainID} 


\author[1,3]{Gonçalo Frutuoso\fnref{fn1}}
\author[2]{Diogo Rodrigues\fnref{fn1}}
\author[1,3]{Alexandre P. Francisco}
\author[2,3]{Cátia Vaz\corref{cor1}}

\cortext[cor1]{Corresponding author: cvaz@cc.isel.ipl.pt}
\fntext[fn1]{These authors contributed equally to this work.}


\affiliation[1]{organization={Instituto Superior Técnico, Universidade de Lisboa}, 
               city={Lisboa}, 
               country={Portugal}}
\affiliation[2]{organization={Instituto Superior de Engenharia de Lisboa, Instituto Politécnico de Lisboa}, 
               city={Lisboa}, 
               country={Portugal}}
\affiliation[3]{organization={Instituto de Engenharia de Sistemas e Computadores, Investigação e Desenvolvimento em Lisboa (INESC-ID)}, 
               city={Lisboa}, 
               country={Portugal}}

\begin{abstract}
Verifying academic credentials remains difficult: records are held by
individual institutions in proprietary systems, verification is slow and manual,
and counterfeit qualifications are widespread. Blockchain-based registries have
been proposed as a remedy, but existing systems tend to anchor certificate
hashes without binding them to a verifiable identity, without an explicit
mechanism to accredit issuing institutions, and without support for correcting
or revoking credentials once issued. This paper investigates whether an
infrastructure designed for regulated financial instruments can be repurposed
to close these gaps. We present the design of a registry for identity-bound
academic credentials that composes OnchainID self-sovereign identities
(ERC-734/ERC-735) with the T-REX suite (ERC-3643): its trusted-issuer registry
becomes an on-chain issuer-accreditation whitelist, and each certificate is
represented as a signed, updatable claim bound to a student's identity and
verifiable by any third party without a wallet, while sensitive fields are kept
off-chain. We make
explicit the tension between a transferable-security-token standard and
non-transferable credentials, clarifying which of its guarantees carry over. We
validate the design with a reference implementation covering the full
certificate life cycle and evaluate it in terms of gas cost, scalability,
latency, and security, quantifying the overhead relative to a hash-anchoring
baseline.

\end{abstract}




\end{frontmatter}

%
%
\section{Introduction}\label{sec:introduction}

The management and verification of academic credentials remain persistent
problems in the education sector. Records are traditionally held by individual
institutions in proprietary systems, so confirming the authenticity of a degree
requires contacting the issuing institution through ad~hoc, often manual,
channels~\cite{OLSON200971}. This opacity has enabled the proliferation of
degree mills~\cite{article} and counterfeit
qualifications~\cite{muzammil2010corrupt}, which erode trust in legitimate
credentials, while the underlying centralized databases remain susceptible to
unauthorized alteration and single points of failure.

Blockchain technology has been widely proposed as a
remedy~\cite{TRIPATHI2023100344, saleh2020blockchain}, and several systems now
record credentials on-chain. Broadly, existing approaches fall into three
families. \emph{Hash-anchoring} systems, such as
Blockcerts~\cite{blockcerts} and the University of Nicosia's Bitcoin-based
diplomas~\cite{unic-pdf}, store the cryptographic digest of a certificate on a
ledger and verify a document by re-computing and comparing its hash.
\emph{Self-sovereign identity} initiatives, most prominently the European
Blockchain Services Infrastructure and its identity framework
ESSIF~\cite{ebsi1, ebsi2}, issue verifiable credentials governed by a
permissioned, institutionally coordinated infrastructure. \emph{Token-based}
schemes, such as EduCTX~\cite{eductx}, represent credits or credentials as
blockchain tokens.

Despite this progress, no existing approach jointly addresses four properties
that a practical academic registry requires. First, a certificate should be
\emph{cryptographically bound to a verifiable identity} rather than existing as
a free-floating document hash. Second, the set of institutions permitted to
issue credentials should be governed by an explicit, on-chain
\emph{accreditation mechanism}, so that authority to issue is itself verifiable.
Third, credentials must be \emph{updatable and revocable}: hash-anchoring
systems are issue-only, yet real registries must correct errors and reflect
post-issuance changes~\cite{guerreiro2022integrating}. Fourth, verification
should be possible for a third party \emph{without requiring a wallet or the
disclosure of sensitive personal data}, in line with data-protection
obligations~\cite{gdprbook, widyasari2024blockchain}. Approaches that emphasise
authenticity, such as Blockcerts and SmartCert~\cite{smartcert}, tend to
overlook identity binding, updatability, and privacy, while top-down frameworks
such as EBSI provide governance but limited flexibility for
correction workflows and non-EU adoption~\cite{ebsi1}.

\paragraph{Approach} This paper investigates whether an infrastructure
originally designed for \emph{regulated financial instruments} can be
repurposed to satisfy these four properties for academic credentials. We build
an academic-certificate registry on two composable Ethereum standards:
OnchainID~\cite{ONCHAINID}, a self-sovereign identity framework in which each
participant is a smart contract holding signed \emph{claims}
(ERC-734/ERC-735~\cite{Vogelsteller_2017_key_manager, Vogelsteller_2017}); and
the T-REX suite~\cite{trexPurposeAndArchitecture}, the reference implementation
of the ERC-3643 permissioned-token standard for compliant security tokens. The
central insight is that the compliance machinery of ERC-3643 maps naturally
onto credentialing: its \emph{trusted-issuer registry} becomes an on-chain
issuer-accreditation whitelist, and its identity-bound, claim-based model
represents each certificate as a signed claim attached to a student's identity
contract. Certificates thus become updatable claims tied to a verifiable
identity, issuable only by accredited institutions, and verifiable by any third
party against the ledger without a wallet, while sensitive fields are kept
off-chain.

Repurposing a security-token standard for credentials also raises a design
tension that we make explicit and examine: ERC-3643 targets \emph{transferable}
regulated assets, whereas academic certificates are inherently
\emph{non-transferable}. We discuss which parts of the standard are exploited
(identity, compliance, issuer accreditation) and which are deliberately
constrained (transferability), and we compare the resulting overhead against a
minimal hash-anchoring baseline to quantify the price of the additional
guarantees. We validate the design with a reference implementation, comprising
the on-chain contracts and a cross-platform mobile client, and evaluate it in
terms of gas cost, scalability, latency, and security.\footnote{The reference
implementation is publicly available at
\url{https://github.com/DiGo-Certify/DiGo-certify-app}.}

\paragraph{Contributions} This paper makes the following contributions:
\begin{itemize}
    \item We present a design for an academic-certificate registry that maps
    the ERC-3643/T-REX compliance model and OnchainID self-sovereign identity
    onto credentialing, using the trusted-issuer registry as an on-chain
    issuer-accreditation mechanism and representing certificates as
    identity-bound, updatable claims
    (Section~\ref{sec:methodology}).
    \item We articulate and analyse the design tension between a
    transferable-security-token standard and non-transferable credentials,
    clarifying which guarantees of ERC-3643 transfer to the credentialing
    setting and which do not (Sections~\ref{sec:methodology}
    and~\ref{sec:evaluation}).
    \item We describe a reference implementation of the complete certificate
    life cycle---institution accreditation, identity creation, issuance,
    update, and wallet-free validation---including a privacy-preserving scheme
    that keeps sensitive fields off-chain (Section~\ref{sec:implementation}).
    \item We provide an empirical evaluation covering per-operation gas cost and
    its monetary cost across deployment scenarios, scalability with respect to
    registry size and claims per identity, latency and throughput, and a
    threat-model-based security analysis, including a quantitative comparison
    with a hash-anchoring baseline (Section~\ref{sec:evaluation}).
\end{itemize}

The remainder of the paper is organised as follows.
Section~\ref{sec:background} reviews the necessary background on blockchain and
credentialing and positions our work with respect to existing approaches.
Section~\ref{sec:methodology} presents the design of the registry and its
mapping onto OnchainID and T-REX. Section~\ref{sec:implementation} describes the
reference implementation. Section~\ref{sec:evaluation} reports the evaluation,
and Section~\ref{sec:conclusions} concludes.


%
%
\section{Background and Related Work}\label{sec:background}

This section provides the background needed to follow the proposed design and
positions it with respect to existing work. We keep the treatment of general
blockchain concepts brief and concentrate on the on-chain identity and token
standards on which our approach builds, and on the landscape of blockchain-based
academic credentialing.

\subsection{Blockchain and Smart Contracts}\label{subsec:blockchain}

A blockchain is a distributed, append-only ledger maintained by a network of
peers without a central authority. Records are grouped into blocks, each
cryptographically linked to its predecessor through a hash, so that any
modification of a past record invalidates every subsequent block; agreement on
the ledger state is reached through a consensus
mechanism~\cite{nakamoto2008bitcoin, xiao2020survey}. These properties make the
ledger tamper-evident and independently verifiable, which is precisely the
guarantee sought for academic records that today rely on a single
institution~\cite{OLSON200971}. Public, permissioned, and consortium
deployments trade off openness against control~\cite{paul2021blockchain}; our
work targets an Ethereum-compatible setting~\cite{tual2015ethereum}, which can
be instantiated on the public network, a Layer-2, or a private consortium chain.

Smart contracts are programs executed deterministically by every node on the
Ethereum Virtual Machine. They allow issuance and verification logic to run
without a trusted intermediary, but their state, once written, is effectively
immutable---a property that underpins integrity yet conflicts with
data-protection requirements such as the right to erasure, a tension we address
through off-chain storage of sensitive data (Section~\ref{sec:implementation}).

\subsection{On-Chain Identity and Token Standards}\label{subsec:standards}

Our design composes two Ethereum standards. \emph{OnchainID}~\cite{ONCHAINID,
onchainidPruposalAndArchitecture} is a self-sovereign identity framework in
which every participant---an individual or an organisation---is represented by
an identity smart contract. These contracts implement ERC-734 (key
manager)~\cite{Vogelsteller_2017_key_manager} and ERC-735 (claim
holder)~\cite{Vogelsteller_2017}, which respectively define how cryptographic
keys and signed \emph{claims} are managed. A claim is an assertion about an
identity---for example a name or a qualification---signed by an issuer and
attached to the subject's identity contract; it can be self-attested or
delegated to an authorised issuer~\cite{enwiki:1234201704}.

\emph{T-REX}~\cite{trexPurposeAndArchitecture} is the reference implementation
of ERC-3643, a standard for \emph{permissioned tokens} that represent regulated
assets. Beyond the token itself, the suite provides a compliance layer whose
central component, for our purposes, is a \emph{trusted-issuer registry}: an
on-chain whitelist recording which identities are authorised to issue claims for
which topics. Although ERC-3643 was conceived for transferable security tokens,
its identity-bound, claim-based compliance model is what we repurpose for
credentialing; the mismatch with the \emph{non-transferable} nature of
certificates, which are closer to \emph{soulbound}
tokens~\cite{weyl2022decentralized} and to non-transferability standards such as
ERC-5192~\cite{erc5192}, is examined in
Sections~\ref{sec:methodology} and~\ref{sec:evaluation}.

\subsection{Blockchain-based Academic Credentialing}\label{subsec:related-work}

Existing systems that place academic credentials on-chain fall into three
families.

\paragraph{Hash-anchoring} The most common approach stores only the
cryptographic digest of a certificate on the ledger and verifies a document by
re-computing and comparing its hash. Blockcerts~\cite{blockcerts} and the
University of Nicosia's Bitcoin-based diplomas~\cite{unic-pdf,
saleh2020blockchain} follow this model, as does
SmartCert~\cite{smartcert}, which adds cryptographic signatures for use during
recruitment. These systems establish authenticity effectively but are
issue-only: they provide no mechanism for correcting or revoking a certificate,
do not bind credentials to a verifiable on-chain identity, and typically require
the holder to share a certificate hash with the verifier, which is neither
convenient nor privacy-preserving.

\paragraph{Self-sovereign identity and verifiable credentials} A growing body of
work applies self-sovereign identity and W3C verifiable
credentials~\cite{w3cVCDataModel, w3cVCDataModel2}---whose data model reached
version~2.0 in 2025---to education~\cite{chan2025blockchain}. At the
infrastructural level, the European Blockchain Services Infrastructure (EBSI) and
its identity framework ESSIF~\cite{ebsi1, ebsi2} issue verifiable credentials over
a permissioned, institutionally governed infrastructure, providing strong
interoperability and governance across member states; this direction is
reinforced by the revised eIDAS regulation
(eIDAS~2.0)~\cite{eidas2}, which mandates interoperable European Digital Identity
(EUDI) Wallets for holding and presenting credentials such as diplomas. However,
this top-down, EU-centric model offers limited flexibility for post-issuance
correction workflows and for adoption outside its governance perimeter. The
QualiChain case
study~\cite{guerreiro2022integrating} integrates an academic management system
(FenixEdu) with Ethereum to register diploma hashes; it demonstrates real-world
feasibility and user acceptance but again centres on authenticity rather than
identity binding, updatability, or privacy.

\paragraph{Token-based} EduCTX~\cite{eductx} represents study credits as
blockchain tokens under a unified ECTS-based scheme, standardising credit
management. It does not, however, address certificate issuance, identity
management, or selective disclosure, and verification requires the holder to
supply a multi-signature address and redeem script. Other token- and
smart-contract-based registries have been proposed~\cite{10163764} with similar
scope.

\subsection{Positioning}\label{subsec:positioning}

Table~\ref{tab:related-comparison} contrasts representative systems on the four
properties identified in Section~\ref{sec:introduction}. No prior approach
jointly provides identity binding, an on-chain issuer-accreditation mechanism,
updatable and revocable credentials, and wallet-free verification with
off-chain protection of sensitive data. Our approach attains all four by
combining OnchainID identities with the ERC-3643 trusted-issuer registry, at a
cost we quantify against a hash-anchoring baseline in
Section~\ref{sec:evaluation}.

\begin{table*}[t]
    \centering
    \caption{Comparison of representative blockchain-based credentialing
    approaches. ``Yes''/``No'' denote support or its absence; ``--'' denotes
    partial support or a property out of the approach's scope.}
    \label{tab:related-comparison}
    \begin{tabular}{l|c|c|c|c|c}
        \hline
        Approach & \makecell{Identity \\ binding} & \makecell{On-chain issuer \\ accreditation} & \makecell{Updatable / \\ revocable} & \makecell{Wallet-free \\ verification} & \makecell{Off-chain \\ private data} \\
        \hline
        Blockcerts / UNIC~\cite{blockcerts, unic-pdf} & No  & No  & No  & --  & --  \\
        SmartCert~\cite{smartcert}                    & No  & No  & No  & No  & No  \\
        EBSI / ESSIF~\cite{ebsi1, ebsi2}              & Yes & --  & --  & --  & Yes \\
        QualiChain~\cite{guerreiro2022integrating}    & No  & --  & No  & --  & --  \\
        EduCTX~\cite{eductx}                          & --  & No  & No  & No  & No  \\
        \hline
        \textbf{Our approach}                         & Yes & Yes & Yes & Yes & Yes \\
        \hline
    \end{tabular}
\end{table*}

\subsection{Adoption Challenges}\label{subsec:challenges}

Beyond technical design, deploying blockchain in education faces
well-documented barriers. Public institutions operate under legal and
administrative constraints and legacy systems that make procedural change
costly, and adoption is further slowed by administrative resistance and the need
for specialised expertise and training~\cite{informatics9030064-barriers,
park2021promises, 8792372-blockchain-based}. Data protection is a central
concern: regulations such as the GDPR grant a right to erasure that conflicts
with the immutability of on-chain records~\cite{gdprbook,
widyasari2024blockchain}, motivating hybrid designs that keep personal data
off-chain and store only integrity-preserving references on-chain. Scalability
and cost are likewise practical considerations, as credential verification
volumes can be high~\cite{app12136380-challenges, zheng2017overview}; these
motivate the deployment scenarios and the empirical cost analysis reported in
Section~\ref{sec:evaluation}.


%
%
\section{Registry Design}\label{sec:methodology}

This section presents the design of the proposed academic-certificate registry.
We first outline the overall architecture, then describe the central
contribution---the mapping of the OnchainID and T-REX (ERC-3643) primitives onto
credentialing---and the design decisions that follow from it: how a
transferable-security-token standard is adapted to non-transferable credentials,
how personal data is protected, and how the certificate life cycle is realised.
Implementation details are deferred to Section~\ref{sec:implementation}.

\subsection{Architecture Overview}\label{subsec:architecture}

\begin{figure*}[t]
    \centering
    \includegraphics[width=0.9\textwidth]{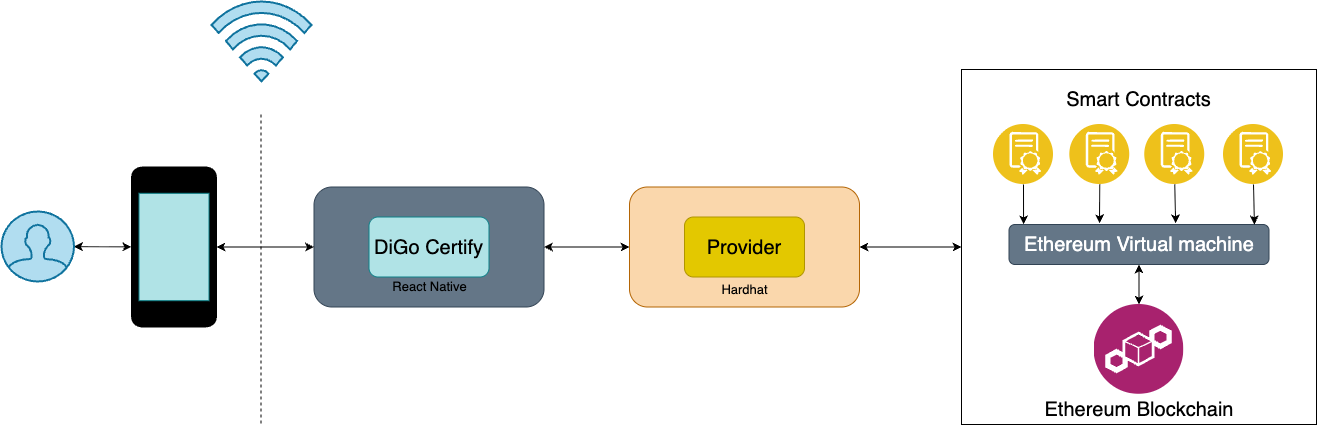}
    \caption{Architecture overview (inspired by~\cite{geeksforgeeks-dApps}).}
    \label{fig:architecture-overview}
\end{figure*}

The system is organised in two layers (Figure~\ref{fig:architecture-overview}).
The \emph{on-chain layer} holds the registry logic and state as smart contracts
on an Ethereum-compatible network; it is where identity, accreditation, and
certificate claims live, and it is the focus of this paper. The \emph{client
layer} is a cross-platform application through which students, institutions, and
verifiers interact with the on-chain layer; its implementation is described in
Section~\ref{sec:implementation}. Confining all trust-bearing logic to the
on-chain layer means that verification does not depend on the client: a third
party can check a certificate directly against the ledger.

\subsection{Mapping Credentialing onto OnchainID and T-REX}\label{subsec:mapping}

The core of the design is the observation that the identity and compliance
primitives of OnchainID and ERC-3643, though conceived for regulated financial
assets, correspond closely to the elements of an academic registry.
Table~\ref{tab:mapping} summarises the correspondence.

\begin{table}[t]
    \centering
    \caption{Mapping of OnchainID/ERC-3643 primitives onto the academic
    registry.}
    \label{tab:mapping}
    \begin{tabular}{p{3.5cm}|p{4.0cm}}
        \hline
        Primitive & Role in the registry \\
        \hline
        OnchainID identity contract (ERC-734/735) & Verifiable on-chain identity of a student or an institution \\
        Signed claim (ERC-735) & A certificate, or a field of it, bound to the student's identity \\
        Claim topic & Category of asserted data (institution, student, certificate) \\
        Trusted-issuer registry (ERC-3643) & On-chain accreditation whitelist of institutions authorised to issue \\
        Claim / issuer key (ERC-734) & Authorisation for an institution to write claims on an identity \\
        \hline
    \end{tabular}
\end{table}

Each participant is represented by an OnchainID identity contract, so that both
students and issuing institutions have a unique, verifiable on-chain
identity~\cite{ONCHAINID, onchainidPruposalAndArchitecture}. A certificate is
modelled not as a standalone document but as a \emph{claim}
(ERC-735~\cite{Vogelsteller_2017})---a statement signed by an issuer and
attached to the subject's identity contract~\cite{enwiki:1234201704}. Composite
certificates are expressed as several claims (e.g.\ name, student number, course,
completion date, registration number, and the certificate digest), each carrying
the issuer's signature.

Authority to issue is itself made verifiable through the ERC-3643 trusted-issuer
registry~\cite{trexPurposeAndArchitecture}, which acts as an on-chain
issuer-accreditation whitelist recording which institutions, and for which claim
topics, are authorised to issue. An
institution's key must additionally be granted on the student's identity
(ERC-734~\cite{Vogelsteller_2017_key_manager}) before it may write claims to it,
so issuance requires both accreditation at the registry level and authorisation
at the identity level. This two-level check is what distinguishes the approach
from hash-anchoring registries, in which any party can record a digest without
its authority being verifiable on-chain. To manage the many identity and issuer
contracts uniformly, the design follows the factory and proxy
patterns~\cite{harmes2008factory, harmes2008proxy} recommended by the
frameworks, the latter allowing the certificate-management logic to be upgraded
without disturbing existing data.

\subsection{Adapting a Security-Token Standard}\label{subsec:transferability}

ERC-3643 targets \emph{transferable} regulated securities, whereas academic
certificates are inherently \emph{non-transferable}, resembling \emph{soulbound}
tokens~\cite{weyl2022decentralized}: a credential is meaningful only when bound
to the individual who earned it. The design therefore exploits
the parts of the standard that concern \emph{identity and compliance}---verified
on-chain identities, the trusted-issuer registry, and the claim-topic model---while
deliberately not using the token's transfer semantics as a means of moving
credentials between holders. In effect, the compliance infrastructure of a
security-token standard is reused as an accreditation-and-identity layer, and the
transferability it was designed to regulate is constrained away. What the
approach gains from ERC-3643 is thus the ready-made, audited machinery for
identity binding and issuer whitelisting; what it does not inherit is a use for
free transfer. We return to this trade-off quantitatively in
Section~\ref{sec:evaluation}, where the cost of this machinery is compared
against a bare hash-anchoring baseline.

\subsection{Privacy by Design}\label{subsec:privacy-design}

Because on-chain state is effectively immutable and publicly readable, personal
data cannot be written to it in the clear without conflicting with
data-protection obligations such as the right to erasure. The design therefore
distinguishes public fields, which may be stored on-chain to allow direct
verification, from sensitive fields, which are kept off-chain; only an
integrity-preserving reference (a salted hash) is recorded on-chain, so that the
ledger attests to a certificate's integrity without exposing its contents. A
symmetric key shared between the student and the issuing institution mediates
access to the private data. Verification of a certificate reduces to comparing a
supplied digest against the on-chain reference and requires neither a wallet nor
the disclosure of the underlying personal data. The concrete storage and
hashing scheme is described in Section~\ref{sec:implementation}.

\subsection{Certificate Life Cycle}\label{subsec:lifecycle}

The registry supports three protocols: issuance, update, and validation.

\begin{figure*}[t]
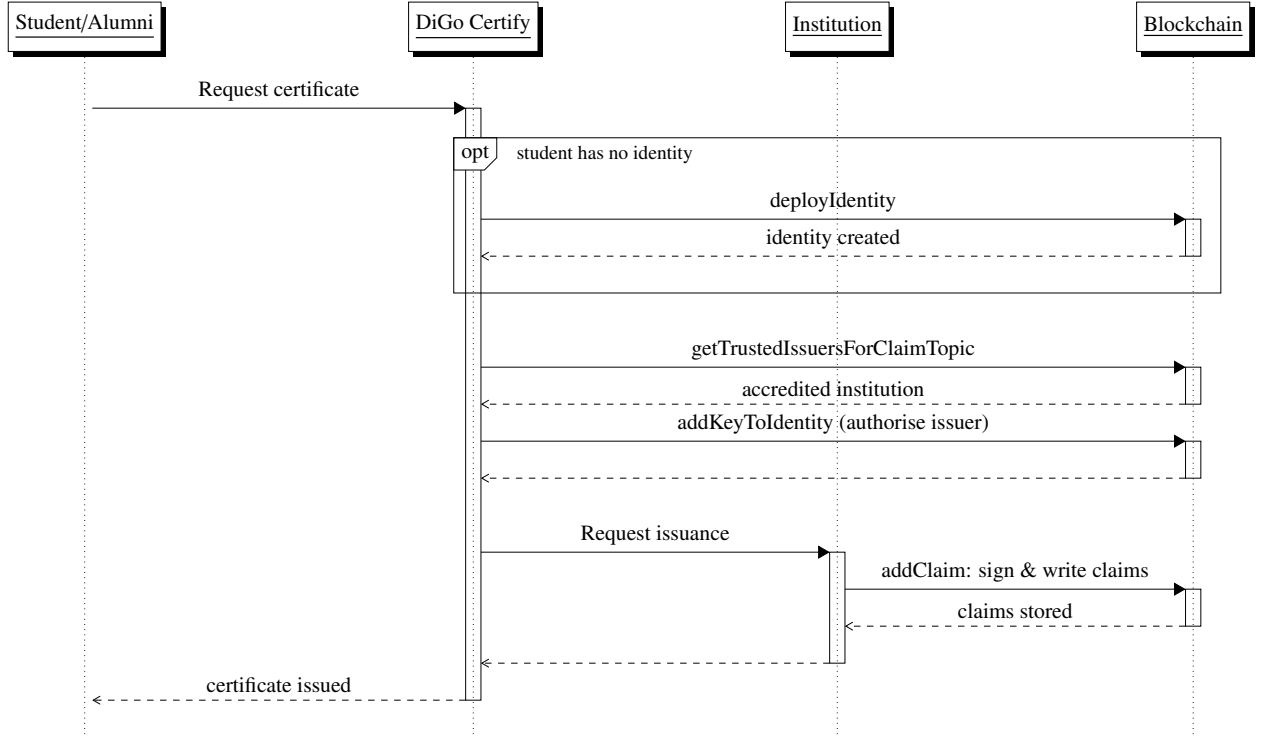

    \centering
    \resizebox{0.92\textwidth}{!}{%
    \begin{sequencediagram}
        \newinst{stu}{Student/Alumni}
        \newinst[4]{app}{DiGo Certify}
        \newinst[4]{inst}{Institution}
        \newinst[4]{bc}{Blockchain}
        \begin{call}{stu}{Request certificate}{app}{certificate issued}
            \begin{sdblock}{opt}{student has no identity}
                \begin{call}{app}{deployIdentity}{bc}{identity created}\end{call}
            \end{sdblock}
            \postlevel
            \begin{call}{app}{getTrustedIssuersForClaimTopic}{bc}{accredited institution}\end{call}
            \begin{call}{app}{addKeyToIdentity (authorise issuer)}{bc}{}\end{call}
            \postlevel
            \begin{call}{app}{Request issuance}{inst}{}
                \begin{call}{inst}{addClaim: sign \& write claims}{bc}{claims stored}\end{call}
            \end{call}
        \end{call}
    \end{sequencediagram}}
    \caption{Issuance of an academic certificate.}
    \label{fig:certificate-request}
\end{figure*}

\paragraph{Issuance} A student requests a certificate from an accredited
institution (Figure~\ref{fig:certificate-request}). After the student's identity
is created and the institution's issuing key is authorised on it, the
institution---having been resolved from the trusted-issuer registry for the
relevant claim topic---signs and writes the certificate claims to the student's
identity. The certificate is thereafter a verifiable, identity-bound record on
the ledger.

\begin{figure*}[t]
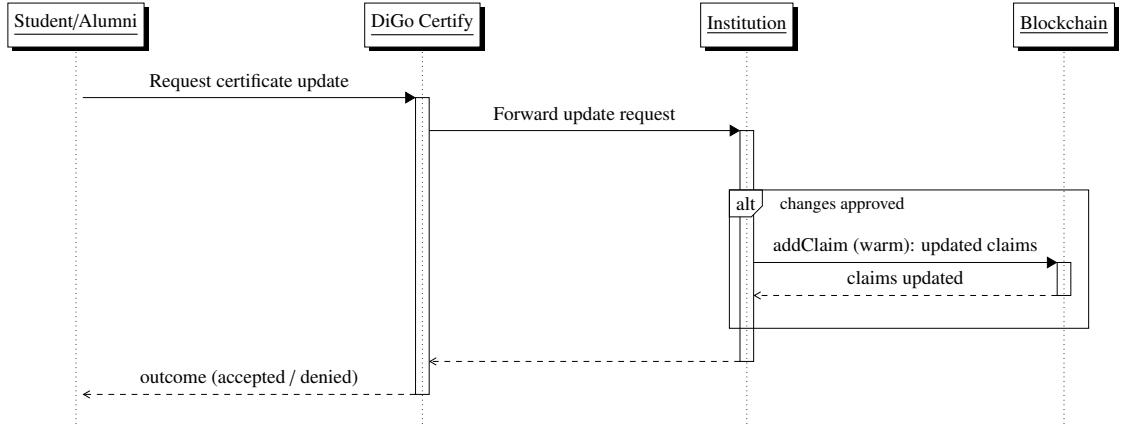

    \centering
    \resizebox{0.82\textwidth}{!}{%
    \begin{sequencediagram}
        \newinst{stu}{Student/Alumni}
        \newinst[4]{app}{DiGo Certify}
        \newinst[4]{inst}{Institution}
        \newinst[4]{bc}{Blockchain}
        \begin{call}{stu}{Request certificate update}{app}{outcome (accepted / denied)}
            \begin{call}{app}{Forward update request}{inst}{}
                \postlevel
                \begin{sdblock}{alt}{changes approved}
                    \begin{call}{inst}{addClaim (warm): updated claims}{bc}{claims updated}\end{call}
                \end{sdblock}
            \end{call}
        \end{call}
    \end{sequencediagram}}
    \caption{Requesting an update to an academic certificate.}
    \label{fig:certificate-update}
\end{figure*}

\paragraph{Update and revocation} Unlike issue-only hash-anchoring systems, the
registry supports post-issuance correction (Figure~\ref{fig:certificate-update}).
A student initiates a change request; upon approval, the accredited institution
issues updated claims, superseding the previous ones. Because claims are managed
by authorised keys, corrections and revocations remain gated by the same
accreditation checks as issuance, and every change is recorded on the ledger.

\begin{figure*}[t]
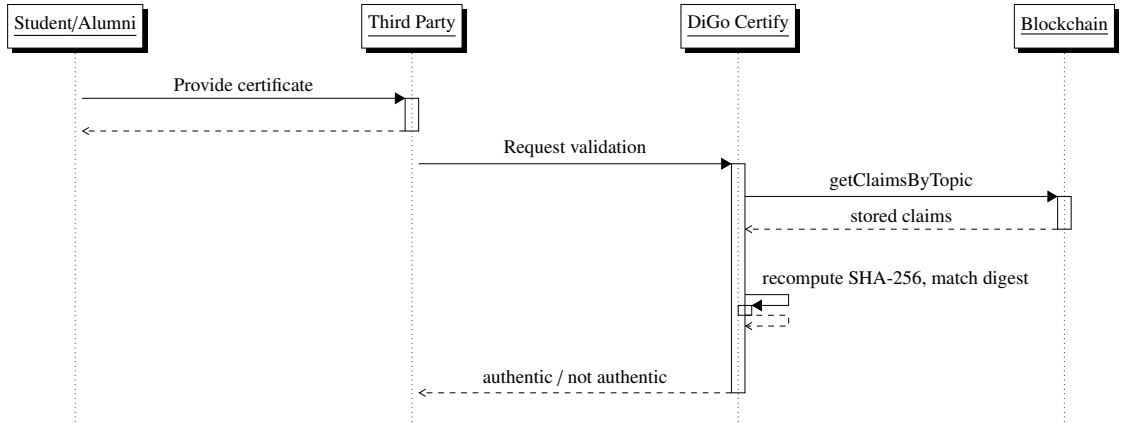

    \centering
    \resizebox{0.82\textwidth}{!}{%
    \begin{sequencediagram}
        \newinst{stu}{Student/Alumni}
        \newinst[4]{tp}{Third Party}
        \newinst[4]{app}{DiGo Certify}
        \newinst[4]{bc}{Blockchain}
        \begin{call}{stu}{Provide certificate}{tp}{}\end{call}
        \begin{call}{tp}{Request validation}{app}{authentic / not authentic}
            \begin{call}{app}{getClaimsByTopic}{bc}{stored claims}\end{call}
            \postlevel
            \begin{call}{app}{recompute SHA-256, match digest}{app}{}\end{call}
            \postlevel
        \end{call}
    \end{sequencediagram}}
    \caption{Validation of an academic certificate.}
    \label{fig:certificate-validation}
\end{figure*}

\paragraph{Validation} A third party verifies a certificate by checking the
supplied digest against the claim stored on the student's identity
(Figure~\ref{fig:certificate-validation}). The check is a read-only operation
against the ledger: it requires no wallet, exposes no private data, and does not
rely on the issuing institution being reachable, allowing any verifier to
independently confirm authenticity.

\subsection{Security Considerations}\label{subsec:design-security}

The registry's guarantees rest on three mechanisms established above: the
tamper-evidence and immutability of the underlying ledger and its
EVM-based execution~\cite{EVM, Wood2014}; the two-level gating of issuance by the
trusted-issuer registry and per-identity key authorisation, which prevents
unaccredited parties from writing valid claims; and the signing of every claim by
its issuer, which makes forgery detectable. Residual risks---such as issuer-key
compromise, denial of service through registry growth, and privacy leakage via
on-chain references---are analysed under an explicit threat model, together with
the corresponding mitigations, in Section~\ref{sec:evaluation}.


%
%
\section{Reference Implementation}\label{sec:implementation}

We validate the design with a reference implementation comprising the on-chain
contracts and a cross-platform client that realises the complete certificate
life cycle. This section describes how the design of
Section~\ref{sec:methodology} is realised in code; the implementation is
released as open
source.\footnote{\url{https://github.com/DiGo-Certify/DiGo-certify-app}} We focus
on the elements that embody the
design---the claim-based data model, the accreditation and issuance paths, and
the privacy-preserving storage and validation scheme---rather than on the user
interface.

\subsection{Overview}\label{subsec:impl-overview}

The on-chain layer reuses two contract suites. OnchainID identities are deployed
and managed through the \texttt{@onchain-id/identity-sdk} package, using its
\texttt{IdentityFactory} to instantiate identity contracts uniformly. The T-REX
(ERC-3643) suite, for which no equivalent package is provided, is deployed from
imported contracts by a dedicated script. All contracts are compiled and
deployed with Hardhat~\cite{hardhat} (Solidity~0.8.17/0.8.20, optimizer enabled,
\texttt{runs}~$=$~200); the deployment emits each contract's address and ABI to
a configuration file consumed by the client. The client targets a single network
through the RPC endpoint recorded in this configuration file; retargeting it to
another chain (e.g.\ a consortium or Layer-2 network) amounts to regenerating the
configuration against the new endpoint and redeploying the contract suite there,
rather than switching networks at run time. The client is a React
Native~\cite{reactNativeSite, ReactNativeBook}/Expo~\cite{Expo} application in
JavaScript~\cite{Javascript} that interacts with the chain through
ethers.js~\cite{ethers, ethersDocs}; wallet keys are held in the device secure
enclave via Expo Secure Store~\cite{Expo-Secure-Store}, and wallet connections
use WalletConnect~\cite{Wallet-Connect}. Contract deployment and the generation
of the configuration file are automated by an installation script accessible
only to the registry owner, so that the creation of accredited issuers remains a
privileged operation.

\subsection{Claim-based Certificate Model}\label{subsec:impl-claims}

Certificates are represented as ERC-735 claims under three topics---%
\texttt{INSTITUTION}, \texttt{STUDENT}, and \texttt{CERTIFICATE}---each carrying
a subset of the certificate fields (for example, the institution's official code
and course identifier, the student's name and number, and the certificate
reference). A claim records its topic (the \texttt{keccak256} digest of the
topic name), a signature scheme (ECDSA), the issuer identity, the issuer's
signature, the claim data, and a \texttt{uri} field; its identifier is derived
as \texttt{keccak256(issuer, topic)}. Each claim is signed by the issuing
institution over the tuple \texttt{(identity, topic, data)}, so that any party
can later verify both the content and its origin. This realises directly the
identity-bound, signed-claim model of Section~\ref{subsec:mapping}.

\subsection{Issuer Accreditation and Identity Creation}\label{subsec:impl-accreditation}

Institutions are onboarded as \emph{claim issuers}. The \texttt{deployClaimIssuer}
routine deploys a claim-issuer contract for an institution's wallet and records
it in the trusted-issuer registry through \texttt{addTrustedIssuer}, together
with the set of claim topics it is permitted to issue; the institution then
registers its official code for later lookup. This is the on-chain
issuer-accreditation step, and, being performed by the installation script, is
restricted to the registry owner.

A student's identity is created by \texttt{deployIdentity}, which deploys an
identity contract for the student's wallet using a unique salt and confirms
creation by observing the corresponding on-chain event. Before an institution
may write claims to a student's identity, its key is authorised on that identity
via \texttt{addKeyToIdentity}, implementing the second, per-identity level of the
two-level gating described in Section~\ref{subsec:mapping}.

\subsection{Issuance}\label{subsec:impl-issuance}

Claims are added through \texttt{addClaim}. At the application layer this first
verifies that the issuer is present in the trusted-issuer registry (or is the
identity's self-signer), rejecting the request otherwise; on-chain, the identity
contract independently requires that the sender hold an authorised claim key
(\texttt{onlyClaimKey}) and that the issuer's signature be valid
(\texttt{isClaimValid}). Accreditation is thus checked at the client while key
authorisation and signature validity are enforced on-chain, realising the
two-level gating of Section~\ref{subsec:mapping}. To request a certificate, the client resolves the
accredited issuer for the relevant topic
(\texttt{getTrustedIssuersForClaimTopic}) and records self-attested student
claims (name and number). Issuance by an institution proceeds identically, with
the institution signing and adding the remaining claims---institution and course
data, the certificate reference, and the certificate itself. Updating a
certificate reissues the affected claims under the same checks, so corrections
remain gated by accreditation.

\subsection{Privacy-Preserving Storage}\label{subsec:impl-privacy}

The implementation separates public from sensitive data, as required by the
design of Section~\ref{subsec:privacy-design}. Public fields are stored on-chain
in the clear to permit direct verification. Sensitive fields are encrypted with
AES-256 under a symmetric key shared between the student and the issuing
institution before being placed in the claim, so that only the intended parties
can read them. The certificate document itself is kept off-chain; what is
anchored on-chain is its SHA-256 digest, stored in the claim's \texttt{uri}
field. The ledger therefore attests to a certificate's integrity without
exposing its contents, keeping personal data off-chain while preserving
verifiability.

\subsection{Validation}\label{subsec:impl-validation}

Validation requires neither a wallet nor the issuing institution to be online. A
verifier presents the certificate (as a link or its contents); the client
retrieves the certificate claims of the student's identity
(\texttt{getClaimsByTopic}) and recomputes the SHA-256 digest of the presented
certificate, reporting the certificate as authentic when the digest matches the
\texttt{uri} of a stored claim. Because this is a read-only comparison against
on-chain state, any third party can independently confirm authenticity without
holding a wallet or accessing the private claim data.


%
%
%
\section{Evaluation}\label{sec:evaluation}

This section evaluates the proposed academic-certificate registry built on
OnchainID and the T-REX (ERC-3643) suite, described in
Sections~\ref{sec:methodology} and~\ref{sec:implementation}. Rather than
assessing the mobile client, we focus on the on-chain layer, which determines
the feasibility, cost, scalability, and security of the approach. The
evaluation is organised around five research questions:

\begin{itemize}
    \item[\textbf{RQ1}] \textbf{Feasibility.} What is the gas cost of each
    operation, and the resulting monetary cost under realistic deployment
    scenarios?
    \item[\textbf{RQ2}] \textbf{Overhead.} How much more expensive is the
    identity- and compliance-based design than a bare hash-anchoring baseline,
    and what does the additional cost provide?
    \item[\textbf{RQ3}] \textbf{Scalability.} How do issuance and validation
    costs grow as the trusted-issuer registry and the number of claims per
    identity increase?
    \item[\textbf{RQ4}] \textbf{Performance.} What end-to-end latency and
    throughput can the system sustain?
    \item[\textbf{RQ5}] \textbf{Security.} Against a defined threat model,
    which attacks are mitigated, and what residual risk remains?
\end{itemize}

\subsection{Experimental Setup}\label{subsec:eval-setup}

We evaluate the system on three environments, each targeting a different class
of metric. Gas consumption is deterministic for a given bytecode, input, and
state, and is therefore measured once on a local development network; monetary
cost is then modelled across chains and gas prices, latency is measured on a
live network, and throughput is bounded analytically from the block gas limit.

\begin{itemize}
    \item \textbf{Local development network.} A Hardhat~\cite{hardhat} node,
    used to obtain deterministic gas measurements (RQ1--RQ2) and to run the
    scalability sweeps (RQ3).
    \item \textbf{Public testnet.} The Ethereum Sepolia testnet, used to
    measure end-to-end latency and to confirm that gas usage matches the local
    network (RQ4).
\end{itemize}

A Proof-of-Authority consortium network (e.g.\ Clique) is arguably the realistic
inter-institutional target deployment; we bound its throughput analytically below
and leave sustained profiling under a tuned block interval to future work.

All measurements were obtained with Solidity 0.8.17/0.8.20, optimizer enabled
with \texttt{runs}~$=$~200, and each compiler's default EVM target (no explicit
\texttt{evmVersion} was set); the exact toolchain versions and the commit hash of
the evaluated code are reported in the public repository.\footnote{The
measurement harness and raw data are available at
\url{https://github.com/DiGo-Certify/DiGo-certify-app}.} Each recurring
state-changing operation was executed over
$R = 30$ independent runs using fresh inputs (a new wallet and salt per run);
we report the median together with the minimum and maximum, and deterministic
one-time deployments were sampled over three runs.
Where an operation writes to previously unset storage, we report the first-write
(\emph{cold}) and rewrite (\emph{warm}) cost separately, as the two differ
substantially at the EVM level.

\subsection{Cost Analysis (RQ1)}\label{subsec:eval-cost}

We first measure the one-time deployment cost of the contracts that constitute
the registry. Table~\ref{tab:eval-deploy} reports the gas consumed and the
deployed bytecode size of each contract, the latter compared against the
EIP-170 limit of 24{,}576 bytes. The two OnchainID identity contracts dominate:
the \texttt{ClaimIssuer} and \texttt{Identity} implementations reach 81\% and
71\% of the code-size limit, respectively, whereas the T-REX registries and the
token remain below 60\%. Their size is the reason both are deployed behind
proxies, and it foreshadows the onboarding asymmetry discussed below: an
institution's \texttt{ClaimIssuer} is deployed as a full contract, whereas a
per-student identity is instantiated as a lightweight proxy through the factory.

\begin{table*}[t]
    \centering
    \caption{Deployment cost and runtime bytecode size of the implementation
    contracts, reported against the EIP-170 limit (24{,}576 bytes). These are
    one-time costs, amortised across all users; in the live suite they are
    deployed behind proxies, and per-student identities are proxies created
    through the factory at 456{,}619 gas (Table~\ref{tab:eval-ops}). The total
    excludes the proxies, the implementation authority, and the factory.}
    \label{tab:eval-deploy}
    \begin{tabular}{l|c|c|c}
        \hline
        Contract & \makecell{Deployment \\ gas} & \makecell{Bytecode \\ (bytes)} & \makecell{\% of \\ EIP-170} \\
        \hline
        Token (ERC-3643)                & 3{,}133{,}123 & 14{,}245 & 58.0 \\
        IdentityRegistry                & 1{,}545{,}748 & 6{,}910  & 28.1 \\
        IdentityRegistryStorage         & 1{,}033{,}016 & 4{,}534  & 18.4 \\
        TrustedIssuersRegistry          & 1{,}176{,}500 & 5{,}201  & 21.2 \\
        ClaimTopicsRegistry             & 482{,}260     & 1{,}987  & 8.1  \\
        ModularCompliance               & 1{,}266{,}889 & 5{,}619  & 22.9 \\
        Identity (implementation)       & 3{,}837{,}535 & 17{,}383 & 70.7 \\
        ClaimIssuer (implementation)    & 4{,}533{,}082 & 19{,}987 & 81.3 \\
        \hline
        \textbf{Implementation total (one-time)} & 17{,}008{,}153 & -- & -- \\
        \hline
    \end{tabular}
\end{table*}

Table~\ref{tab:eval-ops} reports the gas cost of the recurring operations that
make up the certificate life cycle. The two composite rows---issuing and
updating a certificate---aggregate the underlying calls and represent the
figures of practical interest. Certificate validation is performed off-chain by
recomputing the certificate digest and comparing it against the stored claim, so
it incurs no gas for the caller; its latency is analysed in
Section~\ref{subsec:eval-scalability}. Adding a claim to a previously unset
\texttt{(issuer, topic)} slot (\emph{cold}) costs 355{,}409 gas, whereas
rewriting an existing claim (\emph{warm}), which is also the certificate-update
path, costs only 122{,}105 gas---a $2.9\times$ reduction that reflects the
difference between initialising and updating storage at the EVM level.

\begin{table}[t]
    \centering
    \caption{Gas cost of recurring certificate life-cycle operations (medians on
    the local network over 30 runs). Cold and warm denote first-write and rewrite
    storage costs, respectively. The composite rows are sums of their
    components: issuance authorises the issuer once
    (\texttt{addKeyToIdentity}) and writes the certificate claim
    (\texttt{addClaim} cold), while an update rewrites that claim
    (\texttt{addClaim} warm). The identity-proxy onboarding row is the
    lightweight alternative to \texttt{deployClaimIssuer}
    (\texttt{createIdentity}${+}$claim-key${+}$\texttt{addTrustedIssuer}),
    discussed in the text.}
    \label{tab:eval-ops}
    \begin{tabular}{l|c|c}
        \hline
        Operation & \makecell{Gas \\ (cold)} & \makecell{Gas \\ (warm)} \\
        \hline
        \texttt{deployIdentity} (student)              & 456{,}619 & n/a \\
        \texttt{deployClaimIssuer} (institution)       & 4{,}881{,}061 & n/a \\
        Onboard institution (identity proxy)           & 815{,}083 & n/a \\
        \texttt{addTrustedIssuer}                      & 240{,}611 & n/a \\
        \texttt{addKeyToIdentity}                      & 163{,}518 & n/a \\
        \texttt{addClaim} (single claim)               & 355{,}409 & 122{,}105 \\
        \hline
        \textbf{Issue certificate} (composite)         & 518{,}927 & n/a \\
        \textbf{Update certificate} (composite)        & n/a & 122{,}105 \\
        \hline
        Validate certificate (off-chain read)         & \multicolumn{2}{c}{0 (read-only)} \\
        \hline
    \end{tabular}
\end{table}

Two asymmetries stand out. First, onboarding an institution is an order of
magnitude more expensive than onboarding a student: \texttt{deployClaimIssuer}
consumes 4{,}881{,}061 gas---$10.7\times$ the 456{,}619 of
\texttt{deployIdentity}---because each claim issuer is deployed as a full
contract, whereas student identities are instantiated cheaply through the
factory and proxy pattern. Onboarding the institution as a lightweight identity
proxy instead removes most of this cost (815{,}083 gas; Table~\ref{tab:eval-ops}),
as we detail below.
Second, issuing a certificate for the first time from a given institution
(authorising its key, then writing the certificate claim) costs 518{,}927 gas,
but every subsequent certificate from the same institution costs only a single
claim (355{,}409 gas, or 122{,}105 for an update), since the authorisation
persists on the student's identity.

These figures raise a natural question: why inherit the footprint of a
security-token standard---the \texttt{ClaimIssuer} and \texttt{Identity}
implementations alone reach 81\% and 71\% of the EIP-170 limit---rather than
implement a minimal contract that provides identity binding and issuer
accreditation directly? The choice is deliberate reuse rather than economy of
bytecode. OnchainID and T-REX are widely deployed, externally audited
implementations of the ERC-734/735 and ERC-3643 standards; building on them
yields a claim-and-accreditation model that is interoperable with the broader
ERC-3643 ecosystem and inherits its audited access-control logic, at the price of
carrying token machinery the credential setting never exercises. A bespoke
contract would shed that machinery and lower deployment cost, but at the cost of
a new, unaudited trust base and no standard interoperability---a trade-off we
make explicit here and revisit in Section~\ref{sec:conclusions}.

The main run-time cost this bloat imposes is the one-time, per-institution
\texttt{deployClaimIssuer} (4.9M~gas), which deploys a full \texttt{ClaimIssuer}
contract for each institution. A lightweight alternative is to onboard an
institution exactly as a student is---as a factory-created identity proxy---and
simply record that identity in the trusted-issuer registry: because
\texttt{isClaimValid} (signature recovery plus the claim-key purpose check) is
already provided by the base \texttt{Identity} contract, such a proxy can sign
and be accredited without deploying any per-institution bytecode. Measured
directly in the harness, this alternative---a \texttt{createIdentity}, a
claim-key authorisation on the issuer's identity, and an
\texttt{addTrustedIssuer}---onboards an institution for a median of
815{,}083~gas, a $6.0\times$ reduction from the 4{,}881{,}061 of
\texttt{deployClaimIssuer}, at the price of forgoing the
\texttt{ClaimIssuer}-specific \emph{signature-based revocation} discussed in
Section~\ref{subsec:eval-security}. Because the present system
revokes through \texttt{removeClaim} and \texttt{removeTrustedIssuer} rather than
signature revocation, the two designs are functionally equivalent for our life
cycle; the implementation supports both onboarding paths, so the lightweight one
is available wherever its lower cost outweighs the loss of issuer-side signature
revocation---and recovering that revocation is the direction we leave as future
work (Section~\ref{subsec:eval-security}). Independently, for
budget-constrained public institutions a consortium or Layer-2 deployment, where
this gas carries little or no monetary cost (Table~\ref{tab:eval-scenarios}),
already removes the practical friction.

Because gas price and the fiat exchange rate are volatile, we report monetary
cost as a function of the deployment scenario rather than as a single value.
Table~\ref{tab:eval-scenarios} converts the measured gas of the core operations
into fiat cost across a public-chain gas-price range (10--100\,gwei), a
representative Ether-denominated Layer-2 rollup (0.1\,gwei), and a private
consortium chain on which gas carries no monetary cost. All figures assume an
Ether price of \$3{,}000; the gas price and this rate are illustrative, and the
ordering across scenarios is rate-independent.
On the public chain, the recurring operations are inexpensive---issuing a
certificate costs roughly \$16--156 and onboarding a student \$14--137---while
the one-time onboarding of an institution is the only costly step
(\$146--1{,}464), for the structural reason noted above---though the lightweight
identity-proxy alternative cuts this to \$24--245. On a Layer-2 every
operation costs at most a few cents, and on a private consortium chain gas
carries no monetary cost. A mainnet deployment is therefore an upper bound on
cost, and a consortium or Layer-2 deployment is the economically realistic
target.

\begin{table*}[t]
    \centering
    \caption{Modelled monetary cost (USD) of the measured operations across
    deployment scenarios, from the gas of Table~\ref{tab:eval-ops} at an assumed
    Ether price of \$3{,}000. Mainnet columns assume the indicated gas price; the
    Layer-2 column assumes 0.1\,gwei; the private chain incurs only operational
    cost.}
    \label{tab:eval-scenarios}
    \begin{tabular}{l|c|c|c|c|c}
        \hline
        Operation & \makecell{Mainnet \\ 10\,gwei} & \makecell{Mainnet \\ 30\,gwei} & \makecell{Mainnet \\ 100\,gwei} & \makecell{Layer-2 \\ 0.1\,gwei} & \makecell{Private \\ chain} \\
        \hline
        Onboard institution (\texttt{deployClaimIssuer}) & 146.43 & 439.30 & 1{,}464.32 & 1.46 & 0 \\
        Onboard institution (identity proxy)             & 24.45  & 73.36  & 244.52     & 0.24 & 0 \\
        Onboard student (\texttt{deployIdentity})        & 13.70  & 41.10  & 136.99     & 0.14 & 0 \\
        Issue certificate (composite)                    & 15.57  & 46.70  & 155.68     & 0.16 & 0 \\
        Update certificate (warm)                        & 3.66   & 10.99  & 36.63      & 0.04 & 0 \\
        Validate certificate (off-chain)                 & 0      & 0      & 0          & 0    & 0 \\
        \hline
    \end{tabular}
\end{table*}

\subsection{Comparison with a Hash-Anchoring Baseline (RQ2)}\label{subsec:eval-baseline}

To quantify the overhead of the identity- and compliance-based design, we
compare it against a minimal hash-anchoring baseline in the style of
Blockcerts~\cite{blockcerts}, in which the SHA-256 digest of a certificate is
recorded in a registry contract and verified by lookup.
Table~\ref{tab:eval-baseline} contrasts the two designs on cost and on the
functional properties they provide. The comparison isolates the price paid for
identity binding, issuer accreditation, updatable credentials, and wallet-free
verification.

\begin{table}[t]
    \centering
    \caption{Our approach versus a hash-anchoring baseline. Gas figures are
    per operation; feature rows indicate whether the property is supported.}
    \label{tab:eval-baseline}
    \begin{tabular}{l|c|c}
        \hline
        Metric / Property & \makecell{Hash \\ anchoring} & \makecell{Our \\ approach} \\
        \hline
        Gas to issue a certificate      & 45{,}542 & 518{,}927 \\
        Gas to validate                 & 0 & 0 \\
        \hline
        Identity binding                & No  & Yes \\
        Issuer accreditation            & No  & Yes \\
        Updatable / revocable           & No  & Yes \\
        Wallet-free verification        & Yes & Yes \\
        Selective disclosure of fields  & No  & Partial \\
        \hline
    \end{tabular}
\end{table}

The baseline reduces issuance to a single storage write of a digest, and is
correspondingly cheap---45{,}542 gas per certificate---but it anchors an
opaque hash that is bound neither to the subject who earned the credential nor to
an accredited issuer: any account may record any digest, and there is no gated
correction or revocation. Our approach issues an identity-bound, issuer-gated
certificate for 518{,}927 gas (the composite of authorising the issuer's key and
writing the certificate claim, excluding the amortised one-time onboarding of the
identity and the issuer). Both designs validate off-chain at no gas cost and
without a wallet. The identity and compliance layer therefore costs roughly an
order of magnitude in issuance gas (about $11\times$ the baseline), buying in
return on-chain identity binding, verifiable accreditation, and updatable
credentials that the bare baseline cannot provide---a one-time-per-certificate
overhead that is constant and, at typical layer-2 or consortium gas prices,
negligible in absolute terms.

\subsection{Scalability (RQ3)}\label{subsec:eval-scalability}

We assessed how cost evolves along two axes: the size of the trusted-issuer
registry and the number of claims held by a single identity. Measurements were
taken on the local network, where gas is deterministic; each configuration was
sampled over $R = 10$ repetitions, and we report medians.

\begin{figure}[t]
    \centering
    \includegraphics[width=0.85\linewidth]{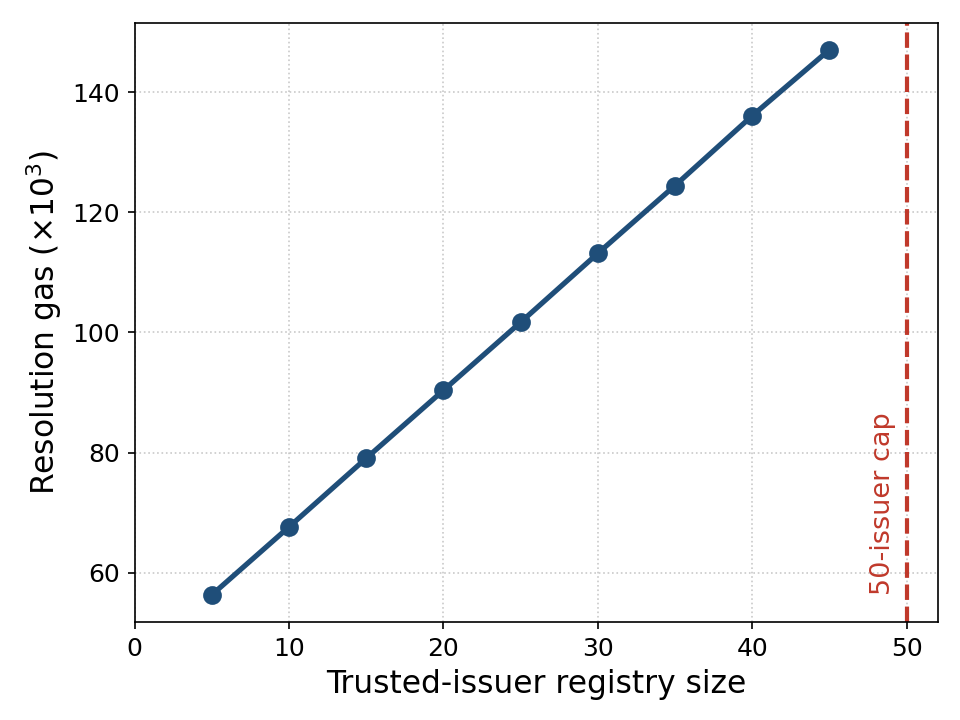}
    \caption{Issuer-resolution gas as a function of trusted-issuer registry
    size: linear ($\approx$2{,}268 gas/issuer), bounded by the registry's
    50-issuer limit.}
    \label{fig:eval-scale-issuers}
\end{figure}

\paragraph{Trusted-issuer registry} Resolving the set of accredited issuers for
a claim topic (\texttt{getTrustedIssuersForClaimTopic}) is on the verification
path, so we grew the registry from 5 to 45 issuers and measured the execution
gas of this resolution (Figure~\ref{fig:eval-scale-issuers}). The cost is linear
in the number of registered institutions, rising from 56{,}337 gas at 5 issuers
to 147{,}053 at 45---a slope of approximately 2{,}268 gas per issuer, i.e.\
resolution gas $\approx 4.5\times10^{4} + 2.27\times10^{3}\,n$. This linear term
is, however, bounded: the T-REX \texttt{TrustedIssuersRegistry} rejects any
addition beyond 50 issuers (\texttt{require(\_trustedIssuers.length < 50)}), so
resolution never exceeds about 150{,}000 gas. The off-chain read latency of the
same call was negligible throughout (median below 4\,ms). The registry is thus
inexpensive but hard-capped at 49 accredited institutions; larger federations
would require partitioning the registry (for instance, one per topic or per
region), which we leave as future work.

\begin{figure}[t]
    \centering
    \includegraphics[width=0.85\linewidth]{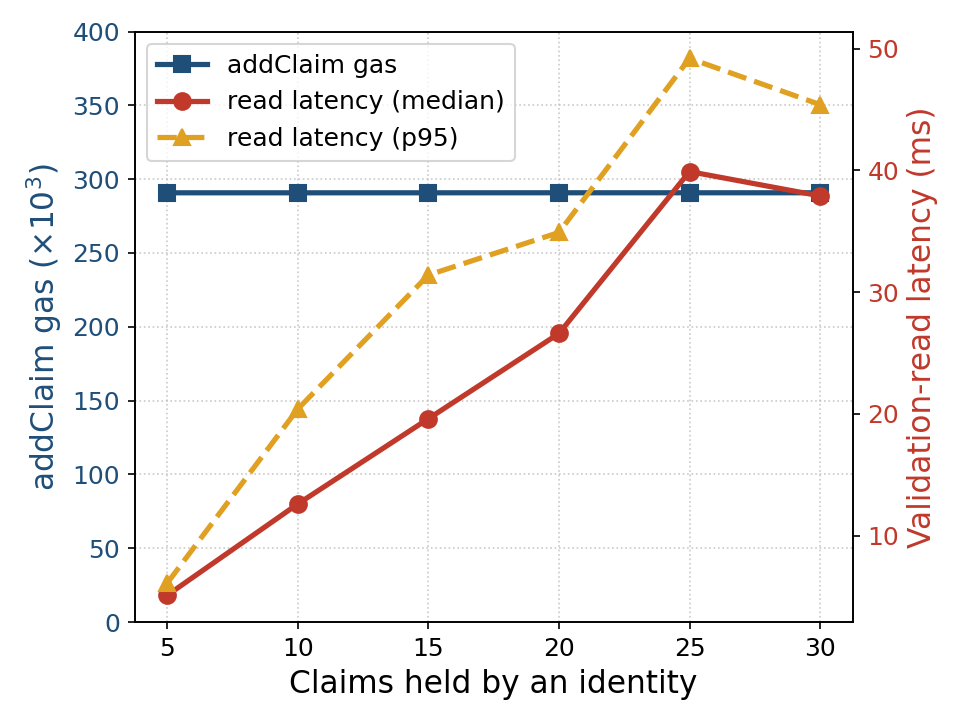}
    \caption{Claim-addition gas (constant) and validation-read latency
    (linear in the number of claims held by an identity).}
    \label{fig:eval-scale-claims}
\end{figure}

\paragraph{Claims per identity} We grew a single identity from 5 to 30
certificate claims---each necessarily issued by a distinct claim issuer, since a
claim is keyed by \texttt{keccak256(issuer,\,topic)}---and measured both the gas
of adding a further claim and the latency of the validation read
(Figure~\ref{fig:eval-scale-claims}). Claim addition is effectively constant at
290{,}756 gas (varying by fewer than 40 gas across all sizes), confirming that
issuance cost is independent of how many claims an identity already holds. (This
absolute figure is lower than the 355{,}409 headline cold cost of
Table~\ref{tab:eval-ops} because the sweep uses a minimal per-claim payload and a
distinct issuer per claim; the invariance, not the absolute value, is the result
of interest here.) The
validation read, in contrast, grows roughly linearly, from a median of 5.1\,ms
at 5 claims to 37.9\,ms at 30 (about 1.3\,ms per claim; 95th percentile up to
49\,ms). This growth is not on-chain gas---validation is read-only---but client
round-trips: retrieving the claims of a topic issues one call to enumerate the
claim identifiers followed by one call per claim. It is therefore
network-dependent, and the reported figures, obtained on an in-process node, are
a lower bound; the cost can be flattened by batching the per-claim reads through
a multicall, an optimisation we leave as future work. In practice a certificate
comprises only a handful of claims, so validation latency remains well within
interactive limits.

\subsection{Latency and Throughput (RQ4)}\label{subsec:eval-perf}

Table~\ref{tab:eval-latency} reports end-to-end latency for issuance and
validation, measured on the Sepolia public testnet over $R = 30$ trials. For
issuance we measure the interval between transaction submission and first
confirmation; for validation we measure the round-trip time of the read-only
call that retrieves an identity's certificate claims for a topic. Because
confirmation times on public networks are heavy-tailed, we report the median and
the 95th percentile.

\begin{table}[t]
    \centering
    \caption{End-to-end latency on the Sepolia public testnet ($R = 30$).
    Issuance is measured from submission to first confirmation; validation is a
    read-only call and is not subject to block confirmation.}
    \label{tab:eval-latency}
    \begin{tabular}{l|c|c}
        \hline
        Operation & Median (s) & p95 (s) \\
        \hline
        Issue certificate    & 11.91 & 24.68 \\
        Validate certificate & 0.10  & 0.16  \\
        \hline
    \end{tabular}
\end{table}

Issuance latency has a median of 11.9\,s, closely tracking Sepolia's
${\approx}12$\,s block interval: the cost is dominated by waiting for block
inclusion rather than by execution. Two ${\approx}25$\,s outliers (24.7 and
25.2\,s) correspond to missed blocks, and over 30 trials they lift the 95th
percentile to 24.7\,s. Validation latency has a median of 0.10\,s (p95
0.16\,s); being a read-only call served by an RPC node, it is independent of
block production and roughly two orders of magnitude faster than issuance. The
gas consumed on Sepolia matched the deterministic local-network figures of
Table~\ref{tab:eval-ops} (a cold first claim of ${\approx}347$k gas and warm
updates of ${\approx}115$k), confirming that gas is chain-independent.

Throughput is bounded by the block gas limit rather than by latency. At a
${\approx}30$M-gas block limit, a cold issuance (${\approx}355$k gas,
Table~\ref{tab:eval-ops}) admits on the order of 85 fresh certificates per block,
and a warm update (${\approx}122$k gas) roughly 245; at Sepolia's ${\approx}12$\,s
interval this is an analytical ceiling of a few to a few tens of issuances per
second. This is a gas-limit ceiling rather than a sustained measurement:
characterising sustained throughput on a private consortium network, where the
block interval is itself a deployment parameter, remains future work.

\subsection{Security Analysis (RQ5)}\label{subsec:eval-security}

We analyse the security of the registry against a threat model covering the
principal assets---certificates, issuer keys, and student identities---and the
trust assumptions of the design. Table~\ref{tab:eval-threats} summarises the
identified threats, the mechanism that mitigates each, whether the mitigation is
exercised by a negative test, and the residual risk. The negative tests are
implemented as an executable suite: each attempts a disallowed action---issuance
by an unaccredited issuer, on-chain issuance without an authorised identity key,
a forged claim signature, use of a revoked claim, issuance by a de-accredited
issuer, a duplicate-salt registration, and exceeding the trusted-issuer
cap---and asserts that it fails. The one threat not covered by a negative test,
on-chain privacy leakage, is a cryptographic property (the dictionary-attack
resistance of a hash of low-entropy data) rather than an access-control check,
and is assessed analytically instead.

\begin{table*}[t]
    \centering
    \caption{Threat model, mitigations, and residual risk. ``Tested'' indicates
    whether a negative test exercises the mitigation; the remaining case (privacy
    leakage) is a cryptographic property assessed analytically rather than by a
    revert test.}
    \label{tab:eval-threats}
    \begin{tabular}{p{2.9cm}|p{4.9cm}|c|p{4.5cm}}
        \hline
        Threat & Mitigation & Tested & Residual risk \\
        \hline
        Unauthorised issuance & On-chain \texttt{onlyClaimKey} and \texttt{isClaimValid}; app-level trusted-issuer check & Yes & Registry owner is trusted; owner-key compromise defeats accreditation \\
        Issuer-key compromise & ERC-734 key rotation (\texttt{removeKey}); \texttt{removeTrustedIssuer} de-accredits & Yes & Claims signed before detection stay valid until the key or issuer is removed \\
        Forged / replayed claim & ECDSA claim signed over \texttt{(identity, topic, data)}; \texttt{isClaimValid} & Yes & No nonce or expiry; a valid signature can be re-added by a claim-key holder \\
        Revocation bypass & \texttt{removeClaim} / \texttt{removeTrustedIssuer}; validation reads current state & Yes & A verifier reading stale state may accept a revoked claim \\
        Registration front-running & Owner-gated \texttt{createIdentity}; salt uniqueness (``salt already taken'') & Yes & Owner manages salts; owner-key leak allows identity squatting \\
        Registry-bloat DoS & Trusted-issuer set capped at 50; only accredited issuers write claims & Yes & A trusted issuer can still bloat an identity, raising read latency \\
        Privacy leakage on-chain & AES-256 on sensitive fields; only the SHA-256 digest stored on-chain & No & Hash of low-entropy data is dictionary-attackable; needs salting \\
        \hline
    \end{tabular}
\end{table*}

Two of the residual risks in Table~\ref{tab:eval-threats}---re-adding a
still-valid signature after a claim has been removed, and a verifier accepting a
revoked claim from stale state---stem from revoking by deletion
(\texttt{removeClaim}) rather than at the issuer. The \texttt{ClaimIssuer}
contract that each institution deploys does, in fact, support \emph{signature-based
revocation} (\texttt{revokeClaimBySignature}), which records a revoked signature
on the issuer so that its \texttt{isClaimValid} rejects the claim thereafter,
independently of the subject's stored state. Our validation path is off-chain and
does not currently consult it, so this capability is presently unused; wiring
validation to check issuer-side revocation---or, more cheaply, adding a nonce or
expiry to each claim---would close the re-add residual. We leave this as future
work, and note that it interacts with the onboarding trade-off of
Section~\ref{subsec:eval-cost}: the lightweight identity-proxy issuer discussed
there forgoes exactly this signature-based revocation, so the two directions form
a deliberate design choice between cheaper onboarding and issuer-side revocation.

In addition to the threat-based analysis, we subjected the compiled contract set
to static analysis using
Slither.\footnote{\url{https://github.com/crytic/slither}} Over the full
compilation (131 contracts) the tool reports 15 high-, 17 medium-, and 75
low-severity findings, alongside informational and optimization notes. Almost all
of these fall within the reused OnchainID and T-REX contracts---for example
reentrancy and unchecked-return-value findings in the token-transfer, delivery,
and modular-compliance machinery---which are externally audited implementations
that our certificate registry does not exercise (it uses only the identity,
claim, and trusted-issuer components, and no token transfers). On the sole
contract we introduce, Slither reports a single informational finding: a
parameter naming-convention deviation with no security impact. We therefore
consider the added attack surface negligible, while noting that reliance on the
upstream contracts inherits their residual findings and that this analysis is not
a formal verification.

Two longer-term privacy considerations fall outside the revert-testable threat
model. First, the encrypted sensitive fields are anchored immutably, so their
confidentiality rests on the continued hardness of AES-256; a deployment with a
long retention horizon should account for eventual cryptographic degradation
(advances in cryptanalysis or large-scale quantum computing) and, accordingly,
keep ciphertext off-chain wherever possible, anchoring only integrity references.
Second, under the GDPR the encrypted personal data is best regarded as
\emph{pseudonymised} rather than anonymous while a decryption key exists, placing
it within scope of the right to erasure (the GDPR ``right to be forgotten''); an
immutable public ledger cannot honour an erasure request directly, whereas a
permissioned consortium deployment, whose participants can collectively prune,
re-key, or destroy the keys for stored state, offers a more defensible legal
posture---reinforcing our conclusion that a consortium network is the
realistic target. Our design mitigates but does not eliminate this exposure,
since only integrity references and encrypted fields, never plaintext personal
data, are placed on-chain.

\subsection{Threats to Validity}\label{subsec:eval-validity}

Gas measurements are deterministic and network-independent, but depend on the
compiler version and optimizer configuration reported in
Section~\ref{subsec:eval-setup}; different settings yield different absolute
values while preserving the relative comparisons. Monetary cost depends on gas
price and exchange rate and is therefore reported as a range rather than a
point estimate. Latency is network-dependent and was measured on a public
testnet (Sepolia); absolute values will differ on other deployments, though the
qualitative separation between validation (read-only) and issuance (write, one
block confirmation) holds. Throughput is reported as an analytical gas-limit
ceiling rather than a sustained measurement. Finally, the
security analysis is limited to the modelled threats and to the coverage of the
static-analysis tools, and does not constitute a formal verification of the
contracts.


%
%
\section{Conclusions}\label{sec:conclusions}

The counterfeiting of academic credentials and the absence of a reliable,
widely accepted system for issuing and verifying them remain significant
problems in the education sector. Existing blockchain-based registries help
establish authenticity but typically anchor certificate hashes without binding
them to a verifiable identity, without an explicit on-chain mechanism to
accredit issuing institutions, and without support for correcting or revoking
credentials once issued.

This paper investigated whether an infrastructure designed for regulated
financial instruments can be repurposed to close these gaps. We presented the
design of a registry for identity-bound academic credentials that composes OnchainID
self-sovereign identities (ERC-734/ERC-735) with the T-REX suite (ERC-3643), in
which the trusted-issuer registry serves as an on-chain issuer-accreditation
whitelist and each certificate is a signed, updatable claim bound
to a student's identity. Issuance is gated at two levels---accreditation in the
registry and per-identity key authorisation---while sensitive data is kept
off-chain and only an integrity-preserving reference is stored on-chain, so that
third parties can verify a certificate without a wallet and without access to
personal data. We made explicit the tension between a
transferable-security-token standard and non-transferable credentials,
identifying which of its guarantees carry over. A reference implementation
covering the full certificate life cycle demonstrates the feasibility of the
approach, and our evaluation characterises its cost, scalability, latency, and
security, including the overhead relative to a hash-anchoring baseline.

Several directions remain for future work. The registry could be deployed and
measured on a public Layer-2 and on a production-grade consortium network to
complement the present evaluation; issuer-side signature-based revocation---%
forgone by the lightweight identity-proxy onboarding that the implementation
already supports (a measured sixfold gas reduction, 815{,}083 versus
4{,}881{,}061~gas)---could be integrated into the validation flow, or replaced by
a per-claim nonce or explicit expiry epoch in the signed claim tuple, to recover
replay-resistant revocation; the privacy model could
be strengthened with
selective-disclosure or zero-knowledge proofs of individual fields; and
integration with recognised credentialing frameworks, such as W3C verifiable
credentials~\cite{w3cVCDataModel2} and the eIDAS~2.0 European Digital Identity
Wallet~\cite{eidas2}, would improve interoperability and support real-world
adoption by institutions.

\section*{Acknowledgements}

This work was supported in part by national funds through FCT ---
Funda\c{c}\~ao para a Ci\^encia e a Tecnologia --- under
projects UID/50021/2025 and UID/PRR/50021/2025 (INESC-ID).

\section*{CRediT author statement}

\textbf{Gonçalo Frutuoso:} Conceptualization, Methodology, Software, Investigation, Writing -- original draft.
\textbf{Diogo Rodrigues:}  Conceptualization, Methodology, Software, Investigation, Writing -- original draft.
\textbf{Alexandre Francisco:} Conceptualization, Computational resources, Supervision, Writing -- review and editing.
\textbf{C\'atia Vaz:} Conceptualization, Supervision, Writing -- original draft, review and editing.

\section*{Code and data availability}

The reference implementation---the on-chain smart contracts, the cross-platform
client, and the measurement harness together with the raw evaluation data---is
publicly available at \url{https://github.com/DiGo-Certify/DiGo-certify-app}.

\section*{Declaration of competing interests}

The authors declare that they have no known competing financial interests
or personal relationships that could have appeared to influence the work
reported in this paper.


\bibliographystyle{elsarticle-num-names}

\bibliography{cas-refs}




\end{document}